\documentclass[12pt,a4paper]{article}
\usepackage{amssymb, amsxtra,txfonts}
\newtheorem{theorem}{Theorem}[section]
\newtheorem{lemma}[theorem]{Lemma}
\newtheorem{proposition}[theorem]{Proposition}
\newtheorem{remark}[theorem]{Remark}

\numberwithin{equation}{section}
\numberwithin{table}{section}
\begin{document}
\begin{center}
{\large Generating Function Associated with the Hankel Determinant Formula for
 the Solutions of the Painlev\'e IV Equation }
\end{center}
\begin{center}
Nalini JOSHI\\ 
School of Mathematics and Statistics F07, The University of Sydney,\\
NSW 2006, Australia\\
nalini@maths.usyd.edu.au\\[3mm]
Kenji KAJIWARA\\
Graduate School of Mathematics, Kyushu University,\\
6-10-1 Hakozaki, Higashi-ku, Fukuoka 812-8512, Japan\\
kaji@math.kyushu-u.ac.jp\\[3mm]
Marta MAZZOCCO\\
School of Mathematics, The University of Manchester,\\
Sackville Street, Manchester M60 1QD, United Kingdom.\\
marta.mazzocco@manchester.ac.uk
\end{center}

\begin{abstract}
We consider a Hankel determinant formula for generic solutions of the
Painlev\'e IV equation.  We show that the generating functions for the
entries of the Hankel determinants are related to the asymptotic
solution at infinity of the isomonodromic problem. Summability of
 these generating functions is also discussed.
\end{abstract}
\begin{center}
\begin{small}
\end{small}
\end{center}

\section{Introduction}
In this article, we consider the Painlev\'e IV equation (P$_{\rm IV}$),
\begin{equation}
 \frac{d^2y}{dt^2}=\frac{1}{2y}\left(\frac{dy}{dt}\right)^2
+\frac{3}{2}y^3 -2ty^2 +\left(\frac{t^2}{2}-\alpha_1+\alpha_0\right)y -
  \frac{\alpha_2^2}{2y},\label{eq:P4}
\end{equation}
where $\alpha_i$ ($i=0,1,2$) are parameters satisfying
$\alpha_0+\alpha_1+\alpha_2=1$. We also denote Equation (\ref{eq:P4}) as
P$_{\rm IV}[\alpha_0,\alpha_1,\alpha_2]$ when it is necessary to specify
the parameters explicitly.  P$_{\rm IV}$ (\ref{eq:P4}) has been studied
extensively {}from various points of view. In particular, it is well-known
that it admits symmetries of affine Weyl group of type $A_2^{(1)}$ as a
group of B\"acklund
transformations\cite{Noumi:book,Noumi-Yamada:P4,Okamoto:P4} (see also
\cite{Luk,AFM,BCH}). Moreover, it is known that P$_{\rm IV}$ admits two
classes of classical solutions for special values of parameters:
transcendental classical solutions which are expressed by the parabolic
cylinder function\cite{Okamoto:P4} and rational solutions
\cite{Gromak,Murata}.  Otherwise the solutions are higher transcendental
functions\cite{Noumi-Okamoto,Umemura-Watanabe}.  Among these solutions,
particular attention has been paid to a class of rational solutions
which are located on the center of the Weyl chambers.  Those solutions
are expressed by logarithmic derivatives of certain characteristic
polynomials with integer coefficients, which are generated by the Toda
equation. Those polynomials are called the Okamoto
polynomials\cite{Noumi:book,Noumi-Yamada:P4,Okamoto:P4,Clarkson}.

The determinant formulas are useful for understanding the nature of the
Okamoto polynomials. In fact, it has been shown that the Okamoto
polynomials are nothing but a specialization of the 3-core Schur
functions by using the Jacobi-Trudi type determinant formula
\cite{Kajiwara-Ohta:P4,Noumi:book,Noumi-Yamada:P4}.  Also, there is
another determinant formula which expresses the Okamoto polynomials in
terms of the Hankel determinant.
Therefore it is an intriguing problem to clarify the underlying meaning
of the Hankel determinant formula. In order to do this, generating
functions for the entries of Hankel determinants have been constructed
in \cite{Goto-Kajiwara}, and it is shown that they are expressed as
logarithmic derivatives of solutions of the Airy equation.

A similar phenomenon has been observed in the study of rational solution
of the Painlev\'e II equation (P$_{\rm
II}$)\cite{Iwasaki-Kajiwara-Nakamura}. Namely, the generating function
associated with the rational solutions of P$_{\rm II}$ is also expressed
as logarithmic derivatives of the Airy function.

Then what do these phenomena mean? In order to answer the question, the
Hankel determinant formula for the {\em generic} solution of P$_{\rm II}$ was
considered in \cite{Joshi-Kajiwara-Mazzocco:P2}. It was shown
that the generating functions of the entries of the Hankel determinant
formula are related to the solutions of isomonodromic problem of
P$_{\rm II}$ \cite{Jimbo-Miwa}. More precisely, the coefficients of
asymptotic expansion of the ratio of solutions of the isomonodromic problem
at infinity give the entries of Hankel determinant formula.
The next natural problem then is to investigate whether such structure can be seen
in other Painlev\'e equations or not.

The purpose of this article is to study the generating functions
associated with the Hankel determinant formula for the generic solutions
of P$_{\rm IV}$ (\ref{eq:P4}). In Section 2, we give a brief review of
the symmetries and $\tau$ functions for P$_{\rm IV}$ through the theory
of the symmetric form of P$_{\rm IV}$\cite{Noumi:book}.  We then construct
the Hankel determinant formula by applying the formula for the Toda
equation\cite{KMNOY:Toda} in Section 3. In Section 4, we consider the 
generating functions for the entries of the Hankel determinant formula. By
linearizing the Riccati equations satisfied by the generating functions,
it is shown that the generating functions are related to the isomonodromic
problem of P$_{\rm IV}$. We also show that the formal series for the generating
functions are summable. Section 5 is devoted to concluding remarks.

\section{Symmetric Form of P$_{\rm IV}$}
In this section we give a brief review of the theory of the symmetric
form of P$_{\rm IV}$. We refer to \cite{Noumi:book}\footnote{The
convention of composition of transformations is different {}from what is
often used in generating complex exact solutions {}from simple ones. See
the section A.4 of \cite{Noumi:book} for details.} for details.
\subsection{Symmetric form and B\"acklund transformations}
The symmetric form of P$_{\rm IV}$ (\ref{eq:P4}) is given by
\begin{equation}
\begin{array}{l}
f_0'=f_0(f_1-f_2)+\alpha_0, \\
f_1'=f_1(f_2-f_0)+\alpha_1,\\
f_2'=f_2(f_0-f_1)+\alpha_2,
\end{array}\label{eq:symmetric_P4}
\end{equation}
where ${}'=d/dt$ and 
\begin{equation}
\alpha_0+\alpha_1+\alpha_2=1,\quad f_0+f_1+f_2=t.
\end{equation}
We obtain P$_{\rm IV}$ (\ref{eq:P4}) for $y=f_2$ by eliminating $f_0$
and $f_1$. The $\tau$ functions $\tau_i$ ($i=0,1,2$) are defined by
\begin{equation}
 h_0=\frac{\tau_0'}{\tau_0},\quad h_1=\frac{\tau_1'}{\tau_1},\quad h_2=\frac{\tau_2'}{\tau_2},
\end{equation}
where $h_i$ ($i=0,1,2$) are Hamiltonians given by
\begin{equation}
\begin{array}{l}
{\displaystyle  h_0=f_0f_1f_2 +
 \frac{\alpha_1-\alpha_2}{3}f_0+\frac{\alpha_1+2\alpha_2}{3}f_1 -
 \frac{2\alpha_1+\alpha_2}{3}f_2,}\\
{\displaystyle  h_1=f_0f_1f_2  -
 \frac{2\alpha_2+\alpha_0}{3}f_0+\frac{\alpha_2-\alpha_0}{3}f_1+\frac{\alpha_2+2\alpha_0}{3}f_2,}
 \\
{\displaystyle  h_2=f_0f_1f_2  +\frac{\alpha_0+2\alpha_1}{3}f_0 -
 \frac{2\alpha_0+\alpha_1}{3}f_1+\frac{\alpha_0-\alpha_1}{3}f_2.}
\end{array}\label{eq:h}
\end{equation}
The symmetric form of P$_{\rm IV}$ (\ref{eq:symmetric_P4}) admits the
following B\"acklund transformations $s_i$ ($i=0,1,2$) and $\pi$ defined
by Table \ref{table:BT}. 
\begin{table}[ht]\label{table:BT}
\begin{displaymath}
 \begin{array}{c|ccc|ccc|ccc}
{} & {\alpha_0} & { \alpha_1} &
{ \alpha_2} & { f_0} & { f_1} &
{ f_2}& { \tau_0}& { \tau_1}& { \tau_2}\\
\hline
{ s_0}&{ -\alpha_0} &{
 \alpha_1+\alpha_0} &{ \alpha_2+\alpha_0} &{
 f_0} &{ f_1+\frac{\alpha_0}{f_0}} &{
 f_2-\frac{\alpha_0}{f_0}} & {
 f_0\frac{\tau_2\tau_1}{\tau_0}} & { \tau_1} & { \tau_2}\\[2mm]
\hline
{ s_1}&{ \alpha_0+\alpha_1} &{
 -\alpha_1} &{ \alpha_2+\alpha_1} &{
 f_0-\frac{\alpha_1}{f_1}} &{ f_1} &{
 f_2+\frac{\alpha_1}{f_1}} &{ \tau_0} & {
 f_1\frac{\tau_0\tau_2}{\tau_1}} &{ \tau_2}\\[2mm]
\hline
{ s_2}&{ \alpha_0+\alpha_2} &{
 \alpha_1+\alpha_2} &{-\alpha_2} &{
 f_0+\frac{\alpha_2}{f_2}} &{ f_1-\frac{\alpha_2}{f_2}} &{
 f_2} &{ \tau_0} & { \tau_1} &{
 f_2\frac{\tau_1\tau_0}{\tau_2}}\\[2mm]
\hline
{ \pi}&{ \alpha_1} &{
 \alpha_2} &{ \alpha_0} &{
 f_1} &{ f_2} &{ f_0} & { \tau_1}
 &{ \tau_2} & { \tau_0}
 \end{array}
\end{displaymath}
\caption{Table of B\"acklund transformations for P$_{\rm IV}$.}
\end{table}
\begin{theorem}
\begin{description}
 \item[(i)] $s_i$ ($i=0,1,2$) and $\pi$ commute with derivation.
 \item[(ii)] $s_i$ ($i=0,1,2$) and $\pi$ satisfy the following fundamental
       relations
\begin{equation}
s_i^2=1,\quad (s_is_{i+1})^3=1,\quad \pi^3=1,\quad \pi
 s_i=s_{i+1}\pi,\quad i\in\mathbb{Z}/3\mathbb{Z},
\end{equation}
and thus $\langle s_0,s_1,s_2,\pi\rangle$ form the extended affine Weyl
group of type $A_2^{(1)}$.
\end{description} 
\end{theorem}
\subsection{Bilinear equations for $\tau$ functions}
The $f$-variables and the $\tau$ functions are related by
\begin{equation}
\begin{array}{l}
\smallskip
{\displaystyle f_0=\frac{\tau_2'}{\tau_2}-\frac{\tau_1'}{\tau_1} +
 \frac{t}{3}=\frac{s_0(\tau_0)\tau_0}{\tau_2\tau_1},}\\
\smallskip
{\displaystyle f_1=\frac{\tau_0'}{\tau_0}-\frac{\tau_2'}{\tau_2} +
 \frac{t}{3}=\frac{s_1(\tau_1)\tau_1}{\tau_0\tau_2},}\\
{\displaystyle f_2=\frac{\tau_1'}{\tau_1}-\frac{\tau_0'}{\tau_0} +
 \frac{t}{3}=\frac{s_2(\tau_2)\tau_2}{\tau_1\tau_0}.}
\end{array}
\label{eq:f-tau}
\end{equation}
Moreover, it is shown that $\tau$ functions satisfy various bilinear
differential equations of Hirota type. For example, we have:

\noindent (I) 
\begin{eqnarray}
&{\displaystyle
 \left(D_t+\frac{t}{3}\right)\tau_2\cdot\tau_1=s_0(\tau_0)\tau_0},\label{bl1:1}\\
& {\displaystyle
 \left(D_t+\frac{t}{3}\right)\tau_0\cdot\tau_2=s_1(\tau_1)\tau_1},\label{bl1:2}\\
& {\displaystyle
 \left(D_t+\frac{t}{3}\right)\tau_1\cdot\tau_0=s_2(\tau_2)\tau_2}.\label{bl1:3}
\end{eqnarray}
\noindent (II) 
\begin{eqnarray}
& {\displaystyle 
\left(D_t^2+\frac{t}{3}D_t-\frac{2}{9}t^2+\frac{\alpha_0-\alpha_1}{3}\right)\tau_0\cdot\tau_1=0,}
\label{bl2:1}\\
& {\displaystyle 
\left(D_t^2+\frac{t}{3}D_t-\frac{2}{9}t^2+\frac{\alpha_1-\alpha_2}{3}\right)\tau_1\cdot\tau_2=0,}
\label{bl2:2}\\
& {\displaystyle 
\left(D_t^2+\frac{t}{3}D_t-\frac{2}{9}t^2+\frac{\alpha_2-\alpha_0}{3}\right)\tau_2\cdot\tau_0=0.}
\label{bl2:3}
\end{eqnarray}
\noindent (III)
\begin{eqnarray}
&& \left(\frac{1}{2}D_t^2 -
    \frac{\alpha_1-\alpha_2}{3}\right)\tau_0\cdot\tau_0=s_1(\tau_1)s_2(\tau_2),\label{bl3:1}\\
&& \left(\frac{1}{2}D_t^2 -
    \frac{\alpha_2-\alpha_0}{3}\right)\tau_1\cdot\tau_1=s_2(\tau_2)s_0(\tau_0),\label{bl3:2}\\
&& \left(\frac{1}{2}D_t^2 -
    \frac{\alpha_0-\alpha_1}{3}\right)\tau_2\cdot\tau_2=s_0(\tau_0)s_1(\tau_1),\label{bl3:3}
\end{eqnarray}
where $D_t^n$ is the Hirota derivative defined by
\begin{equation}
 D_t^n~f\cdot g=\left.\left(\frac{d}{dt}-\frac{d}{ds}\right)^n f(t)g(s)\right|_{s=t}.
\end{equation}
\subsection{Translations and $\tau$ functions on lattice}
Define the translation operators $T_i$ ($i=1,2,3$) by
\begin{equation}
 T_1=\pi s_2s_1,\quad T_2=\pi T_1\pi^{-1}=s_1\pi s_2,\quad
T_3=\pi T_2 \pi^{-1}=s_2s_1\pi.
\end{equation}
Then it follows that 
\begin{equation}
 T_iT_j=T_jT_i\quad(i\neq j),\quad T_1T_2T_3=1.
\end{equation}
Actions of $T_i$ on parameters are given by
\begin{equation}
 \begin{array}{lll}
T_1(\alpha_0)=\alpha_0+1, & T_1(\alpha_1)=\alpha_1-1, & T_1(\alpha_2)=\alpha_2,  \\
T_2(\alpha_0)=\alpha_0,   & T_2(\alpha_1)=\alpha_1+1, & T_2(\alpha_2)=\alpha_2-1,  \\
T_3(\alpha_0)=\alpha_0-1, & T_3(\alpha_1)=\alpha_1,   & T_3(\alpha_2)=\alpha_2+1.  
 \end{array}
\end{equation}
We define 
\begin{equation}
 \tau_{l,m,n}=T_1^lT_2^mT_3^n(\tau_0),
\end{equation}
so that
\begin{equation}
\tau_0=\tau_{0,0,0},\quad \tau_1=\tau_{1,0,0},\quad \tau_2=\tau_{0,0,-1}.
\end{equation}
We note that $\tau_{l+k,m+k,n+k}=\tau_{l,m,n}$ follows {}from
$T_1T_2T_3=1$. Applying $T_1^lT_2^mT_3^n$ on the bilinear equations
(\ref{bl1:1})-(\ref{bl3:3}), we obtain the following equations:

\noindent (I)
\begin{eqnarray}
&&{\displaystyle
 \left(D_t+\frac{t}{3}\right)\tau_{l+1,m+1,n}\cdot\tau_{l+1,m,n}=\tau_{l+1,m,n-1}\tau_{l,m,n}},
\label{bl4:1}\\
&& {\displaystyle
 \left(D_t+\frac{t}{3}\right)\tau_{l,m,n}\cdot\tau_{l+1,m+1,n}=\tau_{l,m+1,n}\tau_{l+1,m,n}},
\label{bl4:2}\\
&& {\displaystyle
 \left(D_t+\frac{t}{3}\right)\tau_{l+1,m,n}\cdot\tau_{l,m,n}=\tau_{l,m-1,n}\tau_{l+1,m+1,n}}.
\label{bl4:3}
\end{eqnarray}
\noindent (II) 
\begin{eqnarray}
&& {\displaystyle 
\left(D_t^2+\frac{t}{3}D_t-\frac{2}{9}t^2+\frac{\alpha_0-\alpha_1+2l-m-n}{3}\right)\tau_{l,m,n}
\cdot\tau_{l+1,m,n}=0,}
\label{bl5:1}\\
&& {\displaystyle 
\left(D_t^2+\frac{t}{3}D_t-\frac{2}{9}t^2+\frac{\alpha_1-\alpha_2-l+2m-n}{3}\right)\tau_{l+1,m,n}
\cdot\tau_{l+1,m+1,n}=0,}
\label{bl5:2}\\
&& {\displaystyle 
\left(D_t^2+\frac{t}{3}D_t-\frac{2}{9}t^2+\frac{\alpha_2-\alpha_0-l-m+2n}{3}\right)\tau_{l+1,m+1,n}
\cdot\tau_{l,m,n}=0.}
\label{bl5:3}
\end{eqnarray}
\noindent (III) 
\begin{small}
\begin{eqnarray}
&& \left(\frac{1}{2}D_t^2 -
  \frac{\alpha_1-\alpha_2-l+2m-n}{3}\right)\tau_{l,m,n}\cdot\tau_{l,m,n}=\tau_{l,m+1,n}\tau_{l,m-1,n},
\label{bl6:1}\\
&& \left(\frac{1}{2}D_t^2 -
  \frac{\alpha_2-\alpha_0-l-m+2n}{3}\right)\tau_{l+1,m,n}\cdot\tau_{l+1,m,n}=\tau_{l+1,m,n+1}\tau_{l+1,m,n-1},
\label{bl6:2}\\
&& \left(\frac{1}{2}D_t^2 -
 \frac{\alpha_0-\alpha_1+2l-m-n}{3}\right)\tau_{l+1,m+1,n}\cdot\tau_{l+1,m+1,n}=\tau_{l+2,m+1,n}\tau_{l,m+1,n}.\label{bl6:3}
\end{eqnarray}
\end{small}
\begin{remark}
Suppose $\tau_0=\tau_{0,0,0}$, $\tau_1=\tau_{1,0,0}$ and
 $\tau_2=\tau_{1,1,0}$ satisfy the bilinear equations
 (\ref{bl2:1})-(\ref{bl2:3}). Then, $f_0$, $f_1$ and $f_2$ defined by
 Equations (\ref{eq:f-tau}) satisfy the symmetric form of P$_{\rm IV}$
       (\ref{eq:symmetric_P4}). This can be verified as
       follows. Dividing Equations (\ref{bl2:1}) and (\ref{bl2:2}) by
       $\tau_0\tau_1$ and $\tau_1\tau_2$, respectively, we have
\begin{eqnarray*}
&&
(h_0+h_1)'+(h_0-h_1)^2+\frac{t}{3}(h_0-h_1)-\frac{2}{9}t^2+\frac{\alpha_0-\alpha_1}{3}=0,\\
&& 
(h_1+h_2)'+(h_1-h_2)^2+\frac{t}{3}(h_1-h_2)-\frac{2}{9}t^2+\frac{\alpha_1-\alpha_2}{3}=0.
\end{eqnarray*}
Subtracting the second equation	{}from the first equation we have
\begin{eqnarray*}
0 &=&  (h_0-h_2)'+(h_0-h_2)(h_0-2h_1+h_2)+\frac{t}{3}(h_0-2h_1+h_2)+\frac{1}{3}-\alpha_1\\
&=& \left(h_0-h_2+\frac{t}{3}\right)'
+ \left(h_0-h_2+\frac{t}{3}\right)(h_0-2h_1+h_2)-\alpha_1=0 \\
&=& f_1'+f_1(f_0-f_2)-\alpha_1,
\end{eqnarray*}
which is the first equation in Equation (\ref{eq:symmetric_P4}). Here we have used the relations
\begin{displaymath}
 h_0-h_2=f_1-\frac{t}{3},\quad  h_1-h_0=f_2-\frac{t}{3},\quad  h_2-h_1=f_2-\frac{t}{3},
\end{displaymath}
which follow {}from Equation (\ref{eq:h}). Other equations in
       Equation  (\ref{eq:symmetric_P4}) can be derived in a similar manner.
\end{remark}
\begin{remark}
Applying $T_1^lT_2^mT_3^n$ on Equation (\ref{eq:f-tau}), we have
\begin{equation}
\begin{array}{l}
{\displaystyle
 T_1^lT_2^mT_3^n(f_0)=\frac{\tau_{l+1,m+1,n}'}{\tau_{l+1,m+1,n}}
-\frac{\tau_{l+1,m,n}'}{\tau_{l+1,m,n}} + \frac{t}{3}
=\frac{\tau_{l+2,m+1,n}\tau_{l,m,n}}{\tau_{l+1,m+1,n}\tau_{l+1,m,n}},}\\
{\displaystyle T_1^lT_2^mT_3^n(f_1)
=\frac{\tau_{l,m,n}'}{\tau_{l,m,n}}-\frac{\tau_{l+1,m+1,n}'}{\tau_{l+1,m+1,n}}
+ \frac{t}{3}
=\frac{\tau_{l,m+1,n}\tau_{l+1,m,n}}{\tau_{l,m,n}\tau_{l+1,m+1,n}},}\\
{\displaystyle T_1^lT_2^mT_3^n(f_2)
=\frac{\tau_{l+1,m,n}'}{\tau_{l+1,m,n}}-\frac{\tau_{l,m,n}'}{\tau_{l,m,n}} +
 \frac{t}{3}=\frac{\tau_{l,m-1,n}\tau_{l+1,m+1,n}}{\tau_{l+1,m,n}\tau_{l,m,n}}.}
\end{array}\label{eq:f-tau:2}
\end{equation}
\end{remark}
\section{Hankel Determinant Formula}
Now consider the sequence of $\tau$ functions $\tau_{n,0,0}$
($n\in\mathbb{Z}$), which are $\tau$ functions in the direction of $T_1$
on the line $\alpha_2={\rm const.}$ in the parameter space. It is
possible to regard this sequence as being generated by the Toda equation
\begin{equation}
\left(\frac{1}{2}D_t^2 -
 \frac{\alpha_0-\alpha_1+2n-1}{3}\right)\tau_{n,0,0}\cdot\tau_{n,0,0}
=\tau_{n+1,0,0}\tau_{n-1,0,0},\label{eq:Toda:tau}
\end{equation}
{}from $\tau_0=\tau_{0,0,0}$ and $\tau_1=\tau_{1,0,0}$. We note that
Equation (\ref{eq:Toda:tau}) follows {}from a specialization of Equation (\ref{bl6:3}).  Let
us introduce the variables $\kappa_n$ ($n\in\mathbb{Z}$) by
\begin{equation}
 \kappa_n={\rm e}^{-\frac{1}{3}nt^2}~\frac{\tau_{n,0,0}}{\tau_{0,0,0}},
\end{equation}
and put
\begin{equation}
\kappa_{-1}=\psi_{-1},\quad \kappa_1=\psi_1,
\end{equation}
where $\psi_{\pm 1}=\psi_{\pm 1}(t)$. 
Then, Equation (\ref{eq:Toda:tau}) can be rewritten as
\begin{equation}
  \frac{1}{2}D_t^2\kappa_n\cdot\kappa_n
=\kappa_{n+1}\kappa_{n-1}-\psi_{-1}\psi_1\kappa_n^2,\quad
\kappa_{-1}=\psi_{-1},\quad \kappa_{0}=1,\quad \kappa_1=\psi_1,\label{eq:Toda:kappa}
\end{equation}
by using the identities
\begin{eqnarray*}
&& D_t(ab)\cdot(cb)=b^2D_ta\cdot c,\\
&& D_t^2(ab)\cdot(cb)=(D_t^2a\cdot c)b^2 + ac(D_t^2b\cdot b),
\end{eqnarray*}
and Equation (\ref{eq:Toda:tau}) with $n=0$. It is known that $\kappa_n$ can
be expressed by a Hankel determinant as follows\cite{KMNOY:Toda}:
\begin{theorem}\label{thm:det}
$\kappa_n$ is given by
\begin{equation}
 \kappa_n=
\left\{
\begin{array}{ll}
\det (a_{i+j-2})_{1\leq i,j\leq n} & n>0\\
1 & n=0\\
\det (b_{i+j-2})_{1\leq i,j\leq |n|} & n<0
\end{array}
\right.
\end{equation}
where the entries are defined by the recurrence relations,
\begin{equation}
 a_n=a_{n-1}' + \psi_{-1}\sum_{k=0}^{n-2}
 a_ka_{n-2-k},\quad a_0=\psi_1,\label{rec:a}
\end{equation}
\begin{equation}
 b_n=b_{n-1}' + \psi_{1}\sum_{k=0}^{n-2}
 b_kb_{n-2-k},\quad b_0=\psi_{-1}.\label{rec:b}
  \end{equation}
\end{theorem}
We note that 
\begin{equation}
 y_{-1}=-\frac{\psi_{-1}'}{\psi_{-1}}+t,\quad y_0=\frac{\psi_{1}'}{\psi_{1}}+t,
\end{equation}
satisfy P$_{\rm IV}[\alpha_0-1,\alpha_1+1,\alpha_2]$ and P$_{\rm
IV}[\alpha_0,\alpha_1,\alpha_2]$, respectively.  Moreover, 
\begin{equation}
 y_n=\frac{\kappa_{n+1}'}{\kappa_{n+1}}-\frac{\kappa_{n}'}{\kappa_{n}}+t,\label{y_and_kappa}
\end{equation}
satisfies P$_{\rm IV}[\alpha_0+n,\alpha_1-n,\alpha_2]$,
\begin{equation}
 \frac{d^2y_n}{dt^2}=\frac{1}{2y_n}\left(\frac{dy_n}{dt}\right)^2
+\frac{3}{2}y_n^3 -2ty_n^2 +\left(\frac{t^2}{2}-\alpha_1+\alpha_0+2n\right)y_n -
  \frac{\alpha_2^2}{2y_n}.\label{eq:P4_n} 
\end{equation}
We also note that the determinant formula for the $\tau$ sequences in
the directions of $T_2$ and $T_3$ are formulated in a similar manner.
\section{Generating Functions and Isomonodromic Problem}
\subsection{Riccati Equations}
For the determinant formula Theorem \ref{thm:det}, let us consider the
generating functions of the entries
\begin{equation}
F_\infty(t,\lambda)=\sum_{n=0}^\infty a_n~\lambda^{-n}, \qquad  
G_\infty(t,\lambda)=\sum_{n=0}^\infty b_n~\lambda^{-n},\label{formal series}
\end{equation} 
where $a_n$ and $b_n$ are characterized by the recurrence relations
(\ref{rec:a}) and (\ref{rec:b}). Multiplying  
Equations (\ref{rec:a}) and (\ref{rec:b}) by $\lambda^{-n}$, and taking summation over $n$, we
obtain the following Riccati equations for the generating functions $F_\infty$ and $G_\infty$.
\begin{proposition}\label{prop:Riccati_t}
 $F_\infty(t,\lambda)$ and $G_\infty(t,\lambda)$ satisfy the Riccati equations
\begin{eqnarray}
&& \lambda\frac{\partial F}{\partial  t}=-\psi_{-1}F^2+\lambda^2F-\lambda^2\psi_1,
\label{Riccati:Ft}\\
&& \lambda\frac{\partial G}{\partial t}=-\psi_{1}G^2+\lambda^2G-\lambda^2\psi_{-1},
\label{Riccati:Gt}
\end{eqnarray}
respectively.
\end{proposition}
Since $F_\infty$ and $G_\infty$ are defined as formal power series at $\lambda=\infty$, it
is convenient to derive differential equations that they satisfy with respect to $\lambda$.
The following auxiliary recurrence relations for $a_n$ and $b_n$ are useful for this purpose.
\begin{lemma}\label{lem:aux}
 $a_n$ and $b_n$ satisfy
\begin{equation}
 \frac{d}{dt}\left[\psi_{-1}a_n-(\psi_{-1}'-t\psi_{-1})a_{n-1}\right]
+ n\psi_{-1}a_{n-1}=0,\label{auxrec:a}
\end{equation}
and
\begin{equation}
 \frac{d}{dt}\left[\psi_{1}b_n-(\psi_{1}'+t\psi_1)b_{n-1}\right]
- n\psi_{1}b_{n-1}=0,\label{auxrec:b}
\end{equation}
respectively.
\end{lemma}
The proof of Lemma \ref{lem:aux} is achieved by tedious but
straight-forward induction by noticing the relations
\begin{eqnarray}
&& \psi_{1}''+t\psi_1'+2\psi_1^2\psi_{-1}+\left(\alpha_0-\alpha_1\right)\psi_1=0,\label{psi1}\\
&& \psi_{-1}''-t\psi_{-1}'+2\psi_1\psi_{-1}^2
+\left(\alpha_0-\alpha_1-2\right)\psi_{-1}=0,\label{psi-1}
\end{eqnarray}
which follow {}from the bilinear equation (\ref{bl5:1}) with
$(l,m,n)=(0,0,0)$ and $(-1,0,0)$, respectively. Lemma \ref{lem:aux} yields the following
linear partial differential equations for $F_\infty$ and $G_\infty$:
\begin{lemma}\label{lem:linear}
$F_\infty(t,\lambda)$ and $G_\infty(t,\lambda)$ satisfy the linear differential equations
\begin{eqnarray}
&& \left(\lambda+t-\frac{\psi_{-1}'}{\psi_{-1}}\right)\frac{\partial F}{\partial t}
-\lambda\frac{\partial F}{\partial \lambda} \nonumber\\
&&\qquad=
-\,\left[\left(\lambda+t\right)\frac{\psi_{-1}'}{\psi_{-1}}-\frac{\psi_{-1}''}{\psi_{-1}}+2\right]F +
\frac{\lambda}{\psi_{-1}}\left(\psi_{-1}\psi_1\right)',\label{linear:FtFlambda}
\\
&& \left(-\lambda+t+\frac{\psi_{1}'}{\psi_{1}}\right)\frac{\partial G}{\partial t}
-\lambda\frac{\partial G}{\partial \lambda}\nonumber\\
&&\qquad=\,-\left[\left(-\lambda+t\right)\frac{\psi_{1}'}{\psi_{1}}+\frac{\psi_{1}''}{\psi_{1}}+2\right]G 
-\frac{\lambda}{\psi_{1}}\left(\psi_{-1}\psi_1\right)',\label{linear:GtGlambda}
\end{eqnarray} 
respectively.
\end{lemma}
Eliminating $t$-derivatives {}from Equations (\ref{Riccati:Ft}) and
(\ref{linear:FtFlambda}), and {}from Equations (\ref{Riccati:Gt}) and
(\ref{linear:GtGlambda}), respectively, we obtain the following Riccati equations
with respect to $\lambda$.
\begin{proposition}\label{prop:Riccati_lambda}
$F_\infty(t,\lambda)$ and $G_\infty(t,\lambda)$ satisfy the following Riccati equations
 \begin{eqnarray}
\label{Riccati:Flambda}&& \lambda^2\frac{\partial F}{\partial \lambda}
=-\left(\lambda+t-\frac{\psi_{-1}'}{\psi_{-1}}\right)\psi_{-1}F^2\nonumber\\
&&\hskip10pt +\lambda\left(\lambda^2+\lambda t+2\psi_1\psi_{-1}+\alpha_0-\alpha_1\right)F
-\lambda^2\left[\left(\lambda+t\right)\psi_{1}+\psi_{1}'\right],\\
\label{Riccati:Glambda}&& \lambda^2\frac{\partial G}{\partial \lambda}
=-\left(-\lambda+t+\frac{\psi_{1}'}{\psi_{1}}\right)\psi_{1}G^2\nonumber\\
&&\hskip10pt -\lambda\left(\lambda^2-\lambda t+2\psi_1\psi_{-1}+\alpha_0-\alpha_1-2\right)G
-\lambda^2\left[\left(-\lambda+t\right)\psi_{-1}-\psi_{-1}'\right],
\end{eqnarray}
respectively.
\end{proposition}
\subsection{Isomonodromic Problem}
The Riccati equations in Proposition \ref{prop:Riccati_t} and
Proposition \ref{prop:Riccati_lambda} can be linearized into second
order linear differential equations by the standard technique, which yields
isomonodromic problems associated with P$_{\rm IV}$.
\begin{theorem}\label{thm:main}
\begin{description}
\item[(i)] It is possible to introduce the functions $Y_1$ and $Y_2$
       consistently as
\begin{eqnarray}
 F_\infty(t,\lambda)&=&\frac{\lambda}{\psi_{-1}}\left(\frac{1}{Y_1}\frac{\partial
			       Y_1}{\partial t} + \frac{\lambda}{2}\right)\nonumber\\
&=&\frac{\lambda^2}{\psi_{-1}}\frac{1}{\lambda+t-\frac{\psi_{-1}'}{\psi_{-1}}}\nonumber\\
&&\times
\left(\frac{1}{Y_1}\frac{\partial Y_1}{\partial \lambda} +
 \frac{\lambda+t}{2}
+\frac{\psi_{-1}\psi_1+\alpha_0-1}{\lambda}+\frac{\alpha_2}{2\lambda}\right),\label{Y1}\\
 Y_2&=&\frac{1}{\psi_{-1}}\left(\frac{\partial Y_1}{\partial t}+\frac{\lambda}{2}Y_1\right).\label{Y2}
\end{eqnarray}
Then the Riccati equations (\ref{Riccati:Ft}) and (\ref{Riccati:Flambda}) are
       linearized to:
\begin{equation}
 \frac{\partial }{\partial
	  \lambda}\left(\begin{array}{c}Y_1\\Y_2\end{array}\right)
=A\left(\begin{array}{c}Y_1\\Y_2\end{array}\right),\quad
 \frac{\partial }{\partial  t}\left(\begin{array}{c}Y_1\\Y_2\end{array}\right)
=B\left(\begin{array}{c}Y_1\\Y_2\end{array}\right),\label{lin:Y}
\end{equation}
\begin{eqnarray}
 A&=&\left(\begin{array}{cc}
-\dfrac{1}{2} & 0\\
0 & \dfrac{1}{2} \end{array}\right)\lambda
+\left(
\begin{array}{cc}
 -\dfrac{t}{2}&u \\
-\dfrac{z+\alpha_1+\alpha_2}{u} & \dfrac{t}{2}
\end{array}
\right)\nonumber\\
&&\qquad\qquad
+\left(
\begin{array}{cc}
 -\left(z+\dfrac{\alpha_2}{2}\right)& uy_{-1}\\
-\dfrac{z(z+\alpha_2)}{uy_{-1}}& z+\dfrac{\alpha_2}{2}
\end{array}
\right)\lambda^{-1},\label{A:Y}\\
B&=&
\left(
\begin{array}{cc}
 -\dfrac{1}{2}&0 \\
0 & \dfrac{1}{2}
\end{array}
\right)\lambda
+
\left(
\begin{array}{cc}
 0& u\\
-\dfrac{z+\alpha_1+\alpha_2}{u} & 0
\end{array}
\right),\label{B:Y}
\end{eqnarray}
where
\begin{equation}
 u=\psi_{-1},\quad
  y_{-1}=-\frac{\psi_{-1}'}{\psi_{-1}}+t,\quad
z=\psi_{-1}\psi_{1}+\alpha_0-1=\psi_{-1}\psi_{1}-\alpha_1-\alpha_2.\label{uyz}
\end{equation}
Conversely, if $Y_1$ and $Y_2$ are the solutions of linear system (\ref{lin:Y})-(\ref{B:Y}), then
\begin{equation}
 F=\lambda~\frac{Y_2}{Y_1},
\end{equation}
satisfies the Riccati equations (\ref{Riccati:Ft}) and
      (\ref{Riccati:Flambda}).
\item[(ii)] It is possible to introduce the functions $Z_1$ and $Z_2$
       consistently as
\begin{eqnarray}
G_\infty(t,\lambda)&=&\frac{\lambda}{\psi_{1}}\left(\frac{1}{Z_1}\frac{\partial
			       Z_1}{\partial t} + \frac{\lambda}{2}\right)\nonumber\\
&=&\frac{\lambda^2}{\psi_{1}}\frac{1}{-\lambda+t+\frac{\psi_{1}'}{\psi_{1}}}\nonumber\\
&&\times\left(\frac{1}{Z_1}\frac{\partial Z_1}{\partial \lambda} 
+ \frac{-\lambda+t}{2}+\frac{\psi_{-1}\psi_1+\alpha_0-1}{s}+\frac{\alpha_2}{2\lambda}\right),\label{Z1} \\
 Z_2&=&\frac{1}{\psi_{1}}\left(\frac{\partial Z_1}{\partial t}+\frac{\lambda}{2}Z_1\right).\label{Z2}
\end{eqnarray}
Then the Riccati equations (\ref{Riccati:Gt}) and (\ref{Riccati:Glambda}) are
linearized to:
\begin{equation}
 \frac{\partial }{\partial  \lambda}\left(\begin{array}{c}Z_1\\Z_2\end{array}\right)
=C\left(\begin{array}{c}Z_1\\ Z_2\end{array}\right),\quad
 \frac{\partial }{\partial  t}\left(\begin{array}{c}Z_1\\ Z_2\end{array}\right)
=D\left(\begin{array}{c}Z_1\\ Z_2\end{array}\right),\label{lin:Z}
\end{equation}
\begin{eqnarray}
 C&=&\left(\begin{array}{cc}
\dfrac{1}{2} & 0\\
0 & -\dfrac{1}{2} \end{array}\right)\lambda
+\left(
\begin{array}{cc}
 -\dfrac{t}{2}& -v \\
\dfrac{w+\alpha_1+\alpha_2}{v} & \dfrac{t}{2}
\end{array}
\right)\\
&&\qquad\qquad
+\left(
\begin{array}{cc}
w+\dfrac{\alpha_2}{2}& vy_0\\
-\dfrac{w(w+\alpha_2)}{vy_0}& -w-\dfrac{\alpha_2}{2}
\end{array}
\right)\lambda^{-1},\label{C:Z}\\
D&=&
\left(
\begin{array}{cc}
 -\dfrac{1}{2}&0 \\
0 & \dfrac{1}{2}
\end{array}
\right)\lambda
+
\left(
\begin{array}{cc}
 0& v\\
-\dfrac{w+\alpha_1+\alpha_2}{v} & 0
\end{array}
\right),\label{D:Z}
\end{eqnarray}
where
\begin{equation}
v=\psi_{1},\quad
y_0=\frac{\psi_{1}'}{\psi_{1}}+t,\quad
w=\psi_{-1}\psi_{1}+\alpha_0-1=\psi_{-1}\psi_{1}-\alpha_1-\alpha_2.\label{vyw}
\end{equation}
Conversely, if $Z_1$ and $Z_2$ are the solutions of linear system  (\ref{lin:Z})-(\ref{D:Z}), then
\begin{equation}
 G=\lambda~\frac{Z_2}{Z_1},
\end{equation}
satisfies the Riccati equations (\ref{Riccati:Gt}) and (\ref{Riccati:Glambda}).
\end{description}
\end{theorem}
\begin{remark}
The linear systems
(\ref{lin:Y})-(\ref{B:Y}) and (\ref{lin:Z})-(\ref{D:Z}) are nothing but
the isomonodromic problem for P$_{\rm
IV}[\alpha_0-1,\alpha_1+1,\alpha_2]$ and P$_{\rm
       IV}[\alpha_0,\alpha_1,\alpha_2]$, respectively\cite{Jimbo-Miwa}.
In fact, compatibility condition of the linear system
(\ref{lin:Y})-(\ref{B:Y})
\begin{equation}
 \frac{\partial A}{\partial t}-\frac{\partial B}{\partial \lambda}+AB-BA=0,
\end{equation}
gives
\begin{eqnarray}
&& \frac{dz}{dt}=\frac{z^2}{y_{-1}}+\left(\frac{\alpha_2}{y_{-1}}-y_{-1}\right)z
 -(\alpha_1+\alpha_2)y_{-1},\label{eqn:z}\\
&& \frac{dy_{-1}}{dt}=2z+y_{-1}^2-ty_{-1}+\alpha_2,\label{eqn:y}\\
&& \frac{du}{dt}=(-y_{-1}+t)u.\label{eqn:u}
\end{eqnarray}
Eliminating $z$, we have P$_{\rm IV}[\alpha_0-1,\alpha_1+1,\alpha_2]$ for $y_{-1}$
\begin{equation}
 y_{-1}''=\frac{(y_{-1}')^2}{2y_{-1}} +
  \frac{3}{2}y_{-1}^3-2ty_{-1}^2+\left(\frac{t^2}{2}-\alpha_1+\alpha_0-2\right)y_{-1}
-\frac{\alpha_2^2}{2y_{-1}}.
\end{equation}
This fact also establishes the consistency of two expressions of
$F_\infty(t,\lambda)$ in terms of $Y_1$ in Equation (\ref{Y1}). A similar remark holds
true for $G_\infty(t,\lambda)$ and $Z_1$.
\end{remark}
\begin{remark}
{}{}From the first equality of Equation (\ref{Y1}), $Y_1$ can be formally
       expressed as
\begin{equation}
 Y_1={\rm const.}\times\exp\left(-\frac{\lambda^2}{4}-\frac{\lambda t}{2}\right)
\lambda^{\alpha_1+\alpha_2/2}\exp\left(-\sum_{n=1}^{\infty}\lambda^{-n}\int\psi_{-1}a_{n-1}dt~\right),
\label{eq:Y1_by_an}
\end{equation}
which coincides with the known asymptotic behavior of $Y_1$ around
       $\lambda\sim\infty$\cite{Jimbo-Miwa}. 
\end{remark}
\subsection{Solutions of Isomonodromic Problems and Determinant Formula} \label{subsec:iso_and_det}
We have investigated the generating functions $F_\infty$ and $G_\infty$
of entries of the Hankel determinant formula and shown that they
formally satisfy the Riccati equations (\ref{Riccati:Ft}),
(\ref{Riccati:Flambda}) and (\ref{Riccati:Gt}), (\ref{Riccati:Glambda}),
respectively, and that those Riccati equations are linearized into
isomonodromic problems (\ref{lin:Y})-(\ref{B:Y}) and
(\ref{lin:Z})-(\ref{D:Z}) for P$_{\rm IV}$. 

Now let us start {}from the isomonodromic problem
(\ref{lin:Y})-(\ref{B:Y}). We have two linearly independent solutions
around $\lambda\sim\infty$, one of which is related to the generating
function $F_\infty$ by $F=\lambda~Y_2/Y_1$. So let us consider another
solution. It is known that the linear system (\ref{lin:Y})-(\ref{B:Y})
admits the following formal solutions\cite{Jimbo-Miwa} around $\lambda\sim\infty$ 
\begin{eqnarray*}
&& \left(\begin{array}{c}Y_1^{(1)} \\ Y_2^{(1)}\end{array}\right)
=\exp\left(-\frac{\lambda^2}{4}-\frac{\lambda t}{2}\right)\lambda^{\alpha_1+\alpha_2/2}
\left[\left(\begin{array}{c}1 \\0 \end{array}\right)
+\left(\begin{array}{c}y_{11}^{(1)} \\y_{21}^{(1)}
       \end{array}\right)\lambda^{-1}
+\cdots\right],\\
&& \left(\begin{array}{c}Y_1^{(2)} \\ Y_2^{(2)}\end{array}\right)
=\exp\left(\frac{\lambda^2}{4}+\frac{\lambda t}{2}\right)\lambda^{-\alpha_1-\alpha_2/2}
\left[\left(\begin{array}{c}0 \\1 \end{array}\right)
+\left(\begin{array}{c}y_{11}^{(2)} \\y_{21}^{(2)}
       \end{array}\right)\lambda^{-1}
+\cdots
\right].
\end{eqnarray*}
These solutions give
\begin{eqnarray*}
&&F^{(1)}(t,\lambda)=\lambda~\frac{Y_2^{(1)}}{Y_1^{(1)}}
=\lambda\times\frac{y_{21}^{(1)}\lambda^{-1}+\cdots}{1+y_{11}^{(1)}\lambda^{-1}+\cdots}
=a_0+a_1\lambda^{-1}+\cdots,\\
&&F^{(2)}(t,\lambda)=\lambda~\frac{Y_2^{(2)}}{Y_1^{(2)}}
=\lambda\times\frac{1+y_{11}^{(1)}\lambda^{-1}+\cdots}{y_{21}^{(1)}\lambda^{-1}+\cdots}
=\lambda^2(c_0+c_1\lambda^{-1}+\cdots),
\end{eqnarray*}
respectively. Theorem \ref{thm:main} states that both
$F^{(1)}(t,\lambda)$ and $F^{(2)}(t,\lambda)$ satisfy the Riccati
equations (\ref{Riccati:Ft}) and (\ref{Riccati:Flambda}). Conversely,
the above two possibilities of power-series solutions for the Riccati
equations are verified directly.
\begin{proposition}\label{prop:leading_order:F}
The Riccati equations (\ref{Riccati:Ft}) and (\ref{Riccati:Flambda})
 admit only the following two kinds of power-series solutions around
 $\lambda\sim \infty$:
\begin{equation}\label{F formal series}
 F^{(1)}(t,\lambda)=\sum_{n=0}^\infty a_n\lambda^{-n},\quad 
 F^{(2)}(t,\lambda)=\lambda^2\sum_{n=0}^\infty c_n\lambda^{-n}.
\end{equation}
\end{proposition}
The proof of Proposition \ref{prop:leading_order:F} is achieved simply by plugging the
power-series solution 
\begin{displaymath}
 F=\lambda^\rho \sum_{n=0}^\infty u_n \lambda^{-n},
\end{displaymath}
into the Riccati equations (\ref{Riccati:Ft}) and
(\ref{Riccati:Flambda}), and investigating the balance of leading terms.
Then we find that $\rho=0,2$. A similar result can be shown for the
Riccati equations (\ref{Riccati:Gt}) and (\ref{Riccati:Glambda}).
\begin{proposition}\label{prop:leading_order:G}
The Riccati equations (\ref{Riccati:Gt}) and (\ref{Riccati:Glambda})
admit only the following two kinds of power-series solutions around
 $\lambda\sim \infty$:
\begin{equation}\label{G formal series}
 G^{(1)}(t,\lambda)=\sum_{n=0}^\infty b_n\lambda^{-n},\quad 
 G^{(2)}(t,\lambda)=\lambda^2\sum_{n=0}^\infty d_n\lambda^{-n}.
\end{equation}
\end{proposition}
It is obvious that $F^{(1)}(t,\lambda)$ and $G^{(1)}(t,\lambda)$ are
nothing but $F_\infty(t,\lambda)$ and $G_\infty(t,\lambda)$,
respectively. Therefore it is an important problem to investigate
$F^{(2)}(t,\lambda)$ and $G^{(2)}(t,\lambda)$. Now we present two
observations regarding this problem. The first observation is that there are
quite simple relations among those functions:
\begin{proposition}\label{prop:F_and_G}
 The following relations hold:
\begin{equation}
 F^{(2)}(t,\lambda)=\frac{\lambda^2}{G^{(1)}(t,-\lambda)},\quad
 G^{(2)}(t,\lambda)=\frac{\lambda^2}{F^{(1)}(t,-\lambda)}.
\end{equation}
\end{proposition}
{\sl Proof.} Substituting $F(t,\lambda)=\lambda^2/g(t,\lambda)$ into
the Riccati equations (\ref{Riccati:Ft}) and (\ref{Riccati:Flambda}), we
obtain Equations (\ref{Riccati:Gt}) and (\ref{Riccati:Glambda}), respectively,
for $G(t,\lambda)=g(t,-\lambda)$. Choosing
$g(t,\lambda)=G^{(1)}(t,\lambda)$, $F(t,\lambda)$ must be
$F^{(2)}(t,\lambda)$, since its leading order is $\lambda^2$. We obtain
the second equation by the similar argument. $\square$

Secondly, $F^{(2)}(t,\lambda)$ and $G^{(2)}(t,\lambda)$ can be also
interpreted as generating functions of the Hankel determinant formula for
P$_{\rm IV}$. Recall that the determinant formula in Theorem
\ref{thm:det} is for the $\tau$ sequence
$\kappa_n={\rm e}^{-nt^2/3}\tau_{n,0,0}/\tau_{0,0,0}$. The following Proposition states
that $F^{(2)}(t,\lambda)$ and $G^{(2)}(t,\lambda)$ correspond to
different normalizations of the $\tau$ sequence:
\begin{proposition}\label{prop:F2G2}
Let 
\begin{eqnarray}
&&F^{(2)}(t,\lambda)=-\frac{\lambda^2}{\psi_{-1}^2}\sum_{n=0}^\infty
 c_n(-\lambda)^{-n}, \label{eq:F2}\\
&&G^{(2)}(t,\lambda)=-\frac{\lambda^2}{\psi_{1}^2}\sum_{n=0}^\infty d_n(-\lambda)^{-n}, \label{eq:G2}
\end{eqnarray}
be formal solutions of the Riccati equations (\ref{Riccati:Ft}),
(\ref{Riccati:Flambda}) and (\ref{Riccati:Gt}), (\ref{Riccati:Glambda}),
respectively.  Then we have the following:
\begin{description}
 \item[(i)] $c_0=-\psi_{-1}$ and $c_1=\psi_{-1}'$. For $n\geq 2$, $c_n$'s are
       characterized by the recursion relation
\begin{equation}
 c_{n+1}=c_n'+\frac{1}{\psi_{-1}}\sum_{k=2}^{n-1}c_kc_{n+1-k},\quad
c_2=\frac{\psi_{-1}''\psi_{-1}-(\psi_{-1}')^2+\psi_{-1}^3\psi_1}{\psi_{-1}}.\label{eq:c_n}
\end{equation}
 \item[(ii)] $d_0=-\psi_{1}$ and $d_1=\psi_{1}'$. For $n\geq 2$, $d_n$'s are
       characterized by the recursion relation
\begin{equation}
d_{n+1}=d_n'+\frac{1}{\psi_{1}}\sum_{k=2}^{n-1}d_kd_{n+1-k},\quad
d_2=\frac{\psi_{1}''\psi_{1}-(\psi_{1}')^2+\psi_{-1}\psi_1^3}{\psi_{1}}.\label{eq:d_n}
\end{equation}
 \item[(iii)] We put
\begin{eqnarray}
&&\sigma_{-n-1}=\det(c_{i+j})_{i,j=1,\ldots,n}\quad (n>0),\quad \sigma_{-1}=1,\\
&&\theta_{n+1}=\det(d_{i+j})_{i,j=1,\ldots,n}\quad (n>0),\quad \theta_{1}=1.
\end{eqnarray}
Then $\sigma_n$ and $\theta_n$ are related to $\tau_{n,0,0}$ as
\begin{eqnarray}
&&
\sigma_n=\frac{\kappa_n}{\kappa_{-1}}={\rm e}^{-\frac{1}{3}(n+1)t^2}
\frac{\tau_{n,0,0}}{\tau_{-1,0,0}}\quad (n<0),\\
&& \theta_n=\frac{\kappa_n}{\kappa_{1}}={\rm e}^{-\frac{1}{3}(n-1)t^2}
\frac{\tau_{n,0,0}}{\tau_{1,0,0}}\quad (n>0).
\end{eqnarray}
\end{description}
\end{proposition}
\noindent{\sl Proof.} (i) and (ii) can be proved easily by substituting
Equations (\ref{eq:F2}) and (\ref{eq:G2}) into the Riccati equations
(\ref{Riccati:Ft}) and (\ref{Riccati:Gt}), respectively, and collecting
the coefficients of powers of $\lambda$. For (iii), we notice that
{}from the Toda equation (\ref{eq:Toda:kappa}) with $n=-1$, namely
\begin{displaymath}
\psi_{-1}''\psi_{-1}-\left(\psi_{-1}'\right)^2=\kappa_{-2}-\psi_{-1}^3\psi_1,
\end{displaymath}
we have
\begin{displaymath}
c_2=\frac{\psi_{-1}''\psi_{-1}-(\psi_{-1}')^2+\psi_{-1}^3\psi_1}{\psi_{-1}}=\frac{\kappa_{-2}}{\kappa_{-1}}
=\sigma_{-2}.
\end{displaymath}
Moreover, the coefficient of the quadratic term in Equation (\ref{eq:c_n}) can be
regarded as
\begin{displaymath}
 \frac{1}{\psi_{-1}} = \frac{\kappa_0}{\kappa_{-1}}=\sigma_{0}.
\end{displaymath}
Applying Theorem \ref{thm:det}, we find that
$\sigma_{-n-1}=\det(c_{i+j})_{i,j=1,\cdots,n}$ $(n>0)$. The statement for
$d_n$ can be shown by a similar argument. $\square$

\subsection{Summability of the Generating Functions}
To study the growth as $n\to\infty$ of the coefficients $a_n(t)$ (or $b_n(t)$) 
in Equation (\ref{formal series}), we use a theorem proved in \cite{hs}. 
\begin{theorem}[Hsieh and Sibuya \cite{hs}, Theorem XIII-8-3]\label{thm:hs}
Consider the following non-linear differential equation in the variable $s$
\begin{equation}
s^{k+1}\frac{dH}{ds}=c(s)H + s\,b(s,H)\label{eq:hs}
\end{equation}
where $k$ is a positive integer, $c(s)$ is holomorphic in the neighbourhood of $s=0$
and $c(0)\not=0$, and $b(s,H)$ is holomorphic in the neighbourhood of $(s, H)=(0,0)$.
Then equation (\ref{eq:hs}) admits one and only one formal solution $H_f(s)$ of the
form $H_f(s)=\sum_{n=1}^\infty a_n\,s^n$. Moreover, $H_f$ is $k$-summable in any
direction $\arg(s)=\vartheta$ except a finite number of values $\vartheta$. Furthermore,
the sum of $H_f(s)$ in the direction $\arg(s)=\vartheta$ is a solution of Equation (\ref{eq:hs}).
\end{theorem}
Equation (\ref{Riccati:Flambda}) can be put into the form (\ref{eq:hs}) by changing variables to
$\lambda=1/s$ and taking $H=F-a_0=F-\psi_1$. We then obtain Equation (\ref{eq:hs}) with $k=2$,
$c(s)\equiv -1$ and
\begin{eqnarray*}
b(s, H)&=&-\,t\,H+\psi_1'+s\Bigl(\psi_{-1}H^2-(\alpha_0-\alpha_1)(H+\psi_{1})\Bigr)\\
&&\qquad\qquad +s^2(t\psi_{-1}-\psi_{-1}')(H+\psi_1)^2
\end{eqnarray*}
Applying Theorem \ref{thm:hs}, we deduce that Equation (\ref{Riccati:Flambda})
admits one and only one formal solution $F_\infty(\lambda)$ of the form $\sum_{n=0}^\infty a_n\lambda^{-n}$. This formal solution is $2$-summable in any direction 
$\arg(\lambda)=\vartheta$ except a finite number of values $\vartheta$ and its sum
in the direction $\arg(\lambda)=\vartheta$ is a solution of Equation (\ref{Riccati:Flambda}).

The definition of $k$-summability implies that $F_\infty(\lambda)$ is of
Gevrey order $2$, namely, for each $t$, there exist positive numbers
$C(t)$ and $K(t)$ such that
\begin{displaymath}
\bigl|a_n(t)\bigr| < C(t) (n!)^{2}K(t)^n,\qquad {\rm for\ all}\ n\ge 1. 
\end{displaymath}
Clearly, one can prove a similar result for the coefficients $b_n$ of the formal solutions $G_\infty$
in Equation (\ref{formal series}), as well as the coefficients $c_n$, $d_n$ of the formal
series in Equations (\ref{F formal series}) and (\ref{G formal series}). For the formal solutions 
$F^{(2)}(t,\lambda)$ (or $G^{(2)}(t,\lambda)$), we need to apply Theorem \ref{thm:hs} 
to a new $H(s)=s^2\,F^{(2)}-c_0$ (or $s^2\,G^{(2)}-d_0$). 

We also note that summability of those series implies that they are
expressible in terms of inverse Laplace transformation of certain
analytic functions. For details, we refer to \cite{b}.

\section{Concluding Remarks}
In this article, we have constructed a Hankel determinant formula for the
$\tau$ sequence of P$_{\rm IV}$ in the direction of translation
$T_1$. Then we have shown that the generating functions of the entries
are closely related to the solutions of isomonodromic problems. More
precisely, coefficients of asymptotic expansion of the ratio of
solutions for isomonodromic problem give the entries of Hankel
determinant formula. Moreover, we have shown that there exist simple but
mysterious relations among those generating functions. We also
discussed the summability of the generating functions. 

Let us finally give some remarks.  Firstly, in this article we have
considered only the $\tau$ sequences in the direction of $T_1$. It is
also possible to consider the directions $T_2$, $T_3$ by
the $\widetilde{W}(A_2^{(1)})$ symmetry of P$_{\rm IV}$.  Secondly, it is
surprising that the results obtained in this article are completely
parallel to the P$_{\rm II}$ case \cite{Joshi-Kajiwara-Mazzocco:P2},
although concrete computations depend on the specific situation for each
case. In particular, it is remarkable that many formulas have exactly
the same form as the P$_{\rm II}$ case, such as Equation (\ref{eq:Y1_by_an})
(except for the dominant exponential factor), or formulas in Section
\ref{subsec:iso_and_det}. This coincidence may imply that (i) the
phenomena observed for P$_{\rm II}$ and P$_{\rm IV}$ could be universal;
at least it can be seen for other Painlev\'e equations, (ii) the
underlying mathematical structure may originate {}from the Toda equation,
rather than the Painlev\'e equations themselves.

These points will be explored in forthcoming articles.

\end{document}